\documentclass[12pt]{iopart}
\usepackage[colorlinks=true,urlcolor=blue,linkcolor=blue]{hyperref}
\usepackage{graphicx}
\relax

\def\be{\begin{equation}}
\def\ee{\end{equation}}
\def\bs{\begin{subequations}}
\def\es{\end{subequations}}
\def\calm{{\cal M}}
\def\dcalm{\delta {\cal M}}

\def\ex{\epsilon}

\newcommand{\qh}{{\hat{q}}}
\newcommand{\Hh}{{\hat{H}}}
\newcommand{\Hpt}{{H_\theta}}
\newcommand{\Hpr}{{H_r}}

\newcommand{\rb}{{\bar r}}
\newcommand{\bb}{{\bar b}}
\newcommand{\tb}{{\bar \tau}}
\newcommand{\Rb}{{\bar R}}

\newcommand{\rt}{\tilde r}

\newcommand{\ttt}{\tilde t}

\def\be{\begin{equation}}
\def\ee{\end{equation}}
\def\bs{\begin{subequations}}
\def\es{\end{subequations}}
\newcommand{\een}{\end{subequations}}
\newcommand{\ben}{\begin{subequations}}
\newcommand{\beq}{\begin{eqalignno}}
\newcommand{\eeq}{\end{eqalignno}}

\def \lta {\mathrel{\vcenter
     {\hbox{$<$}\nointerlineskip\hbox{$\sim$}}}}
\def \gta {\mathrel{\vcenter
     {\hbox{$>$}\nointerlineskip\hbox{$\sim$}}}}

\newcommand\fverb{\setbox\pippobox=\hbox\bgroup\verb}
\newcommand\fverbdo{\egroup\medskip\noindent%
                        \fbox{\unhbox\pippobox}\ }
\newcommand\fverbit{\egroup\item[\fbox{\unhbox\pippobox}]}
\newbox\pippobox

\def \lta {\mathrel{\vcenter
     {\hbox{$<$}\nointerlineskip\hbox{$\sim$}}}}
\def \gta {\mathrel{\vcenter
     {\hbox{$>$}\nointerlineskip\hbox{$\sim$}}}}

\begin{document}

\title{Cosmological Acceleration and Gravitational Collapse}

\author{Pantelis S. Apostolopoulos$^1$\footnote{E-mail: vdfspap8@uib.es}, Nikolaos Brouzakis$^2$, Nikolaos Tetradis$^2$ and Eleftheria Tzavara$^2$}
\address{$^1$Departament de F\'isica, Universitat de les Illes Balears, Cra. Valldemossa Km 7.5, E-07122 Palma de Mallorca, Spain\\
$^2$University of Athens, Department of Physics, University Campus, Zographou 157 84, Athens, Greece}

\begin{abstract}
The acceleration parameter defined through the 
local volume expansion is negative for 
a pressureless, irrotational fluid with positive energy density. 
In the presence of inhomogeneities or anisotropies 
the volume expansion rate results from averaging 
over various directions. 
On the other hand, 
the observation of light from a certain source in the sky provides information
on the expansion 
along the direction to that source.
If there are preferred directions in the underlying geometry one can define 
several expansion parameters.
We provide such definitions for 
the case of the Tolman-Bondi metric.
We then examine the effect of a localized inhomogeneity 
on the surrounding cosmological fluid. Our framework is similar in spirit 
to the model of spherical collapse. 
For an observer in the vicinity of a central overdensity, the 
perceived local evolution is consistent with acceleration in the 
direction towards the center of the overdensity, 
and deceleration perpendicularly to it. 
A negative mass leads to deceleration along the radial
direction, and acceleration perpendicularly to it. 
If the observer is located at the center of an overdensity the null geodesics
are radial. The form of the luminosity distance as a function of the redshift
is consistent with acceleration for a certain range of redshifts.
\end{abstract}
\maketitle

\section{Introduction}
\setcounter{equation}{0}

The cosmological expansion of our Universe seems to have accelerated in the 
recent past. This conclusion is supported 
by the form of the luminosity distance as a function of the redhshift for
distant supernovae \cite{accel1,accel2}. 
In the context of homogeneous cosmology, a recent accelerating phase is also
in agreement with the observed perturbations in the cosmic microwave
background \cite{wmap}.
The mechanism that triggered 
the acceleration has not been identified conclusively. 
The simplest explanation is that the
cosmological constant is non-zero. However, the absence of acceleration at
redshifts $z\gta 1$ implies that the required value of the cosmological 
constant is approximately 120 orders of magnitude smaller than its natural
value in terms of the Planck scale. 

One intriguing fact is that the accelerating phase coincides with the 
period in which inhomogeneities in the matter distribution at length scales
$\lta 10$ Mpc become significant, so that the Universe cannot be approximated
as homogeneous any more 
at these scales. A link between inhomogeneities and cosmological
acceleration has been pursued in various studies. There have been
arguments, based on perturbative estimates, that the 
backreaction of superhorizon
inhomogeneities on the cosmological expansion 
is significant and could cause the acceleration \cite{kolb}. 
However, the validity of this 
effect is questionable \cite{seljak}.

We are interested in the importance
for the problem of cosmological acceleration
of inhomogeneities with sub-horizon characteristic scales today.
Because of the significant 
growth of such inhomogeneities at recent times, a perturbative
treatment may not be sufficient. An exact solution of the Einstein 
equations, even for a simplified geometry, could be more useful in order to
reveal an underlying mechanism. The Tolman-Bondi metric \cite{tb} has
been employed often in this context 
\cite{rasanen}--\cite{tbother}. It has been observed that 
any form of the luminosity distance as a function of the redshift can be 
reproduced with this metric \cite{mustapha}.

A drawback of the standard handling of the Tolman-Bondi metric is
that the radial
coordinate is defined such that the fluid density is initially perceived as
homogeneous. 
The presence of inhomogeneities is introduced through a function that 
determines the local Big Bang time. 
This obscures the intuition on the role 
of large mass concentrations. Morever, the growth of 
perturbations and its effect on the expansion is not obvious. 

In our analysis we choose a gauge such that 
the initial density perturbation is apparent.
We first model the perturbation by matching a
Schwarzschild metric in the interior with 
an exterior Tolman-Bondi metric.  
We show that the presence of a large overdensity modifies the cosmological 
evolution of the surrounding fluid. In particular, accelerating expansion can 
take place along the radial direction and can be observed through the 
redshift of light signals propagating radially. 
On the other hand, the expansion perpendicularly 
to the radial direction is characterized by deceleration.
Underdensities do not induce acceleration in any direction.

As an exotic possibility, 
we also consider the case of a negative mass at
the center. 
A fluid with energy density that can become
negative has already been discussed in the literature. 
In ref. \cite{phantom} a field termed phantom with
negative kinetic energy drives the accelerating expansion. 
Our setup is based on less drastic assumptions. We consider 
vacuum solutions of the Einstein equations that can be interpreted as arising
from objects of negative mass. The relevant metric is the
Schwarzschild one with a negative mass.
We show that a pressureless fluid near the center of such a configuration
can have accelerating expansion perpendicularly to the 
radial direction.

We should mention at this point that 
negative-mass configurations have several features that appear
problematic, and their physical significance
is not guaranteed. 
The negative-mass Schwarzschild solution
has a naked singularity at the center,
so that it violates the cosmic censorship conjecture. 
However, naked singularities
appear often in studies of gravitational collapse for
large classes of initial conditions \cite{joshi}. 
Another issue concerns 
the stability of the negative-mass Schwarzschild solution, which 
has not been proven in full generality. In ref. \cite{gibbons} the linear
stability has been demonstrated, under a family of boundary conditions
at the singularity that have a physical motivation.
On the other hand, runaway solutions of the Einstein equations are known for 
a pair of objects with positive and negative mass \cite{bondi,gibbons}.
We do not offer here a resolution of these problematic
issues. Instead, we point out an interesting feature of the 
negative-mass Schwarzschild solution related to the issue of the 
cosmological acceleration.

Next we consider a homogeneous energy distribution in the central 
region. 
We define an initial condition for the spatial curvature which 
makes our model
very similar to that of spherical collapse \cite{sphc1}. 
A small overdensity grows consistently with the perturbative Jeans analysis, 
until the density fluctuation becomes comparable to the 
average density. Then, the overdensity starts collapsing. 
We study 
the null geodesics that start outside the perturbation and 
lead to an observer in its interior. If the observer is located in the 
center of the spherical perturbation the geodesics follow the 
radial direction. The accelerated expansion of the surrounding fluid
is reflected in their form, which, in turn, determines the luminosity
distance of a light source as a function of the redshift.

We demonstrate that the expected 
luminosity distance is consistent with accelerated expansion for a certain
range of redshifts. 
In this work we do not put emphasis
on the exact numerical consistency with the data. The reason is that
we model the fluid in the central collapsing region as pressureless.
This results in a strong deceleration for small redshifts, which is not 
expected to be present if the cosmological fluid can reach virialization.
Our main point is the identification of a mechanism that 
could drive the observed acceleration. 
This mechanism 
is based on the gravitational force that acts on the fluid surrounding a 
collapsing overdensity and its effect on the expansion rate.

In the following section we discuss the cosmological expansion in inhomogeneous
cosmologies and its dependence on the direction of observation. In section 3
we employ the Tolman-Bondi metric in order to
construct a model for a central inhomogeneity in an asymptotically 
homogeneous Universe.
In section 4 we examine if acceleration can be 
induced by the inhomogeneity along
or perpendicularly to the radial direction. 
In section 5 we study the form of the luminosity distance as a function of the 
redshift for an observer located at the center of an overdensity. In
section 6 we provide a summary of the results and our conclusions.

\section{Acceleration in inhomogeneous cosmology}
\setcounter{equation}{0}

The presence of a center in the configuration that we would like to study
implies that we have to assume a metric appropriate for an inhomogeneous fluid.
Under the assumption of spherical symmetry,
the most general metric for a 
pressureless, inhomogeneous fluid is the  
Tolman-Bondi metric \cite{tb}.
It can be written in the form 
\begin{equation}
ds^{2}=-dt^2+b^2(t,r)dr^2+R^2(t,r)d\Omega^2,
\label{metrictb}
\end{equation}
where $d\Omega^2$ is the metric on a two-sphere. 
The function $b(r,t)$ is given by
\begin{equation}
b^2(t,r)=\frac{R^2_{,r}(t,r)}{1+f(r)},
\label{brttb} \end{equation}
where the subscript denotes differentiation with respect to $r$, and
$f(r)$ is an arbitrary function.
The bulk energy momentum tensor has the form
\begin{equation}
T^A_{~B}={\rm diag} \left(-\rho(t,r),\, 0,\, 0,\, 0  \right).
\label{enmomtb} \end{equation}
The fluid consists of successive shells marked by $r$, whose
local density $\rho$ is time-dependent. 
The function $R(t,r)$ describes the location of the shell marked by $r$
at the time $t$. Through an appropriate rescaling it can be chosen to satisfy
\cite{joshi}
\be
R(0,r)=r.
\label{s0} \ee

The Einstein equations reduce to 
\begin{eqnarray}
R^2_{,t}(t,r)&=&\frac{1}{8\pi M^2}\frac{\calm (r)}{R}+f(r)
\label{tb1} \\
\calm_{,r}(r)&=&4\pi R^2 \rho \, R_{,r},
\label{tb2} \end{eqnarray}
with 
$G=\left( 16 \pi M^2 \right)^{-1}$.
The generalized mass function $\calm(r)$ of the fluid can be chosen 
arbitrarily. It incorporates the
contributions of all shells up to $r$. It determines the energy density
through eq. (\ref{tb2}). Because of energy conservation
$\calm(r)$  
is independent of $t$, while $\rho$ and $R$ depend on both $t$ and $r$.

The volume expansion rate is usually defined through the four-velocity of 
the fluid 
$u^a$ as
\be
3H =u^a_{~;a}=u_{a;b}\, g^{ab}=u_{a;b} \, h^{ab},
\label{theta} \ee
where
\be
h^{ab}=g^{ab}+u^a u^b.
\label{hab} \ee
The Raychaudhuri equation constrains its evolution 
with time. It reads
\be
\dot{H}\equiv H_{,c}\,u^c=-H^2
-\frac{1}{3}\sigma_{ab} \sigma^{ab}+\frac{1}{3}\omega_{ab} \omega^{ab}
-\frac{1}{3}R_{ab}u^a u^b,
\label{ray} \ee
where
$\sigma_{ab}$ and $\omega_{ab}$ are the shear and vorticity 
tensors respectively. 
For an observer comoving with a pressureless 
and irrotational fluid,
the acceleration parameter is 
\be
q=
\frac{\dot{H}}{H^2}+1=
-\frac{1}{H^2}\left( \frac{1}{3} \sigma_{a b} \sigma^{ab}
+\frac{1}{12 M^2}\rho\right).
\label{rayn} \ee
If the local energy density is positive, no acceleration can take place.

In the context of 
inhomogeneous cosmology the definition (\ref{theta}) does not capture the 
variation of the expansion rate in different directions. 
For example, the Tolman-Bondi (\ref{metrictb}) has a preferred direction
(the radial one). 
In such situations we can define a tensor $p^{ab}$ that projects every 
quantity perpendicularly to the preferred space-like direction 
$s^a$ (and of course the time-like vector field $u^a$). This is
\be
p^{ab}=g^{ab}+u^a\,u^b -s^a \, s^b=h^{ab}-s^a \, s^b.
\label{pab} \ee 
For the Tolman-Bondi metric (\ref{metrictb}) we have $s^a=b^{-1}\partial_r$.
We can now define invariant expansion rates parallel and perpendicularly to
$s^a$, according to \cite{moffat}
\begin{eqnarray} 
\Hpr&=&u_{a;b}\, s^a s^b=\frac{b_{,t}}{b}=\frac{R_{,rt}}{R_{,r}}
\label{tpr1} \\
\Hpt&=&\frac{1}{2}u_{a;b}\,p^{ab} =\frac{R_{,t}}{R}, 
\label{tpt1} \end{eqnarray}
so that 
\be
H=\frac{2}{3}\Hpt+\frac{1}{3}\Hpr.
\label{theta12} \ee
From the above, it is obvious that $H$ corresponds to an average of the 
expansion rates in various directions. 
If the acceleration is
defined through $H$, the expansion is always decelerating
for $\rho >0$. This is the conclusion drawn in previous studies
\cite{rasanen,moffat,alnes}.
However, 
it is possible that the expansion may be accelerating is some direction 
even though eq. (\ref{rayn}) always holds.

A definition of the expansion rate that takes into account its directional
dependence is given in ref. \cite{humph}. It is
\be
\Hh=\frac{1}{3}u^a_{~;a}+\sigma_{ab}J^a J^b,
\label{hdir} \ee
where $\sigma_{ab}$ is the shear tensor and $J^a$ a unit vector pointing in
the direction of observation. In the context of the Tolman-Bondi metric, and 
for an observer located away from the center of the configuration, the above
definition gives \cite{humph}
\be
\Hh=\frac{R_{,t}}{R}+\left(
\frac{b_{,t}}{b}-\frac{R_{,t}}{R} 
\right) \cos^2\psi.
\label{hhtb} \ee
The parameter $\psi$ is the angle between the radial direction through 
the observer and the direction of observation. For $\psi=0$ or $\pi$ we have
$\Hh=H_r$, while for $\psi=\pi/2$ or $3\pi/2$ we have $\Hh=H_\theta$.

A definition of the acceleration parameter in a specific direction 
can be given in terms
of the expansion of the luminosity distance $D_L$ of a light source 
in powers of the redshift $z$ of the incoming photons. For small $z$ this is
\be
\hat{q}=\Hh\frac{d^2D_L}{dz^2}-1.
\label{dddz} \ee
In ref. \cite{humph} the
acceleration parameter was calculated in the context of the Tolman-Bondi
metric for an observer located away from the
center of spherical symmetry. The result is \cite{humph}
\begin{eqnarray}
\hat{q}&=& \frac{1}{\hat{H}^2}\left(
C_3+C_4|\cos\psi|+C_5\cos^2\psi +C_6|\cos\psi|^3
\right)+3
\label{qhat} \\
C_3&=&\frac{R_{,tt}}{R}-3\frac{R^2_{,t}}{R^2}
\label{C3} \\
C_4&=&3\sqrt{1+f}\left(\frac{R_{,t}}{R^2}-\frac{R_{,tr}}{R_{,r}R} \right)
\label{C4} \\
C_5&=&3\frac{R_{,t}^2}{R^2}-3\frac{R_{,tr}^2}{R_{,r}^2}
-\frac{R_{,tt}}{R}+\frac{R_{,ttr}}{R_{,r}}
\label{C5} \\
C_6&=&\sqrt{1+f}\left( 
3\frac{R_{,tr}}{RR_{,r}}+\frac{R_{,tr}R_{,rr}}{R_{,r}^3}-3\frac{R_{,t}}{R^2}
-\frac{R_{,trr}}{R_{,r}^2} \right).
\label{C6}
\end{eqnarray}

For $\psi=0$ or $\pi$ and $\psi=\pi/2$ or $3\pi/2$ the acceleration is, 
respectively,
\begin{eqnarray}
q_r&=&\left(\frac{b}{b_{,t}}\right)^2
\left[\frac{b_{,tt}}{b}-\frac{1}{b}\left(\frac{b_{,t}}{b}\right)_{,r}
\right],
\label{qr} \\
q_\theta&=&\left(\frac{R}{R_{,t}}\right)^2
\frac{R_{,tt}}{R}.
\label{qth} \end{eqnarray}
These parameters do not depend solely on local quantities, as opposed to  
the acceleration parameter of eq. (\ref{rayn}).
For example, from eq. (\ref{tb1}) we have
\be
q_\theta=
-\frac{1}{16\pi M^2}\frac{\calm}{R^3}\frac{1}{H_\theta^2}.
\label{accelpe} \ee
The value of $q_\theta$ depends on the total mass function and not just
the local energy density. 
For $\calm>0$ we expect to observe 
deceleration perpendicularly to the radial direction,
while for $\calm < 0$ acceleration.

It must be emphasized that the expressions (\ref{qr}), (\ref{qth})
refer to the effective acceleration deduced from light signals
that originate in the vicinity of an off-center observer.
In this sense, they determine
the character of the local expansion. It is interesting that they do not
predict a dipole component \cite{humph}. On the other hand, light signals that
travel longer distances (e.g. from outside a spherical inhomogeneity to its
center) integrate information on the local expansion along their whole 
trajectory. We shall discuss the form
of the luminosity distance as 
a function of the redshift for this case in a following section.

Finally, 
we point out that in the standard treatment of the Tolman-Bondi metric 
\cite{rasanen}--\cite{tbother} the radial coordinate is taken such that
the initial energy density is perceived as homogeneous. Moreover, the function
$f(r)$ is assumed to be zero in order to provide consistency with a matter
dominated, flat Universe. The inhomogeneity is introduced through a function
$t_b(r)$ that appears in the integration of the Friedmann equation and
determines the local Big Band time.
Our approach is closer to the physical picture, as the overdensities are
apparent from the beginning. Moreover, we put the
emphasis on their effect on the cosmological evolution of the 
surrounding fluid.

\section{Expansion near an inhomogeneity}
\setcounter{equation}{0}

We would like to describe the cosmological evolution around a central
region in which the mass function deviates from the form implied by a
homogeneous distribution of matter. 
We have in mind the general form 
\be
\calm(r)=\calm_0+\dcalm(r),
\label{dmr} \ee
with $\calm_0$ positive or negative.
Even for negative $\calm_0$ we assume that eq. (\ref{tb2}) 
is satisfied with $\rho\geq 0$. 
No negative energy density
appears in the energy-momentum tensor, and the weak energy condition holds.
We now have 
\be
\dcalm_{,r}=4\pi S^2 \rho \, S_{,r}.
\label{dtb2} \ee

The most convenient way to realize our scenario is by matching
two spherically symmetric space-times. We assume that 
in the central region the metric
takes the Schwarzschild form with mass $\calm_0$
\be
ds^2=-A^2(\rt)d\ttt^2+B^2(\rt)d\rt^2+\rt^2d\Omega^2,
\label{schw} \ee
where $A^2(\rt)=B^{-2}(\rt)=1-\calm_0/(8\pi M^2\rt)$.
The Tolman-Bondi metric (\ref{metrictb})
determines the geometry at large distances.
The two regions are separated by a spherical surface with metric
\be
ds^2=-d\tau^2+S^2(\tau)d\Omega^2.
\label{surf} \ee
The physical situation described by our scenario does not involve 
necessarily a 
singularity at the origin. 
For $\calm_0>0$ the metric of eq. (\ref{schw}) may result from a 
compact object with positive energy density and radius smaller than
$S(\tau)$.
For $\calm_0<0$ we are forced to consider eq. (\ref{schw}) as a vacuum
solution in order to avoid a negative energy density.

The matching of (\ref{metrictb}) and (\ref{surf}) leads to 
\begin{eqnarray}
S^2(\tau)&=&R^2\left( t(\tau),r(\tau) \right)
\label{match11} \\
-{\dot{t}}^2(\tau)+
b^2\left(t(\tau),r(\tau) \right) \,\, \dot{r}^2(\tau)&=&-1,
\label{match12} 
\end{eqnarray}
where a dot denotes a derivative with respect to the 
proper time $\tau$ on the surface.
The matching of (\ref{schw}) and (\ref{surf}) gives
\begin{eqnarray}
S^2(\tau)&=&\rt^2(\tau)
\label{match21} \\
-A^2\left(\rt(\tau) \right)\,\, {\dot{\ttt}}^2(\tau)+
B^2\left(\rt(\tau) \right) \,\, \dot{\rt}^2(\tau)
&=& -1.
\label{match22} 
\end{eqnarray}
The absence of shell crossing implies $\dot{r}(\tau)=0$.
The shell that coincides with the matching surface is always the
same. As a result, it maintains a constant value $r(\tau)=r_0$. 
Eq. (\ref{match12}) then gives
$t(\tau)=\tau$, while eq. (\ref{match11}) can be written as
$S(\tau)=R(\tau,r_0)$.

In order to obtain an equation for $S(\tau)$ we need 
the additional assumption that there is no singular
energy density on the matching surface. This implies that
the extrinsic curvature is continuous across the surface. The resulting 
condition reads
\be
S(\tau)\left(\dot{S}^2(\tau)+\frac{1}{B^2(S(\tau))} \right)^{1/2}
=
\frac{R(\tau,r_0)\, R_{,r}(\tau,r_0)}{b(\tau,r_0)}.
\label{extr} \ee
Making use of eq. (\ref{brttb}) we find
\be
\dot{S}^2(\tau)=f(r_0)+\frac{1}{8\pi M^2}\frac{\calm_0}{S(\tau)}.
\label{eqmatch} \ee
This coincides with the equation of motion of the first shell, as derived from
eqs. (\ref{tb1}), (\ref{tb2}).

We can conclude that, in the absence of shell crossing, 
the interior Schwarzschild metric can be matched with the
exterior Tolman-Bondi one on a spherical surface 
that coincides with the location of the first shell $S(\tau,r_0)$. 
The equation of motion for the shells of fluid 
is 
\be
\frac{\dot{R}^2(\tau,r)}{R^2}=\frac{1}{8\pi M^2}\frac{\calm (r)}{R^3}
+\frac{f(r)}{R^2},
\label{eom1a} \ee
where $\calm(r)$ is given by eq. (\ref{dmr})
with $\dcalm(r)\to 0$ for $r\to r_0$. 
The initial condition is given by eq. (\ref{s0}).
The above equation can be considered as 
a generalization of the standard Friedmann equation for
an inhomogeneous cosmological fluid. The integrated mass within a spherical
volume of comoving radius $r$ is given by $\calm(r)$, while the
function $f(r)$ defines a generalized curvature term.

In order for eq. (\ref{tb1}) to have a solution, the two
arbitrary functions $f(r)$ and $\calm(r)$ must satisfy
$f(r)\geq -\calm(r)/(8\pi M^2r)$ for all $r$. 
The function $f(r)$ defines an effective curvature term 
in eq. (\ref{tb1}). 
We can also interpret $f(r)$ as part of the initial radial velocity of the
fluid. This has to be non-zero in the presence of a negative mass for 
a physically meaningful solution to exist.

As a physically motivated example
we consider a fluid with energy density
that is initially constant for $r\geq r_0$, so that 
$\dcalm(r)=4\pi\rho_0\left(r^3-r^3_0 \right)/3$.
We assume that a spherically symmetric inhomogeneity has appeared during
the cosmological evolution of the fluid (probably as a result of gravitational
instability). At $t_i=0$ the inhomogeneity is 
concentrated in the region $r < r_0$. We parametrize the total mass 
of the inhomogeneity as 
$\calm_0={4\pi} \rho r_0^3/3$ with
$\rho=\ex \rho_0$. 
For $\ex>1$ the inhomogeneity is a local overdensity, for $0\leq \ex <1$ an
underdensity, while for $\ex < 0$ we have an object of negative mass at the
center of the configuration.

Our crucial assumption is 
that the expansion rate at some initial time $t_i$
is given for all $r$ by the standard
expression in homogeneous cosmology:
$H_i^2=\rho_0/(6M^2)$.
We essentially assume that the expansion
is completely homogeneous at the time of appearance of the 
inhomogeneity. 
This is very similar to the initial condition considered 
within the context of the 
model of spherical collapse \cite{sphc1, sphc2}.
Then, eq. (\ref{eom1a}) with $R(0,r)=r$ implies that 
$\bar{f}(\rb) =   (1-\ex)/{\rb}$,
where $\bar{f}=f/\bar{H}_i^2$ 
and $\bar{H}_i=H_i r_0 $. 
An effective curvature term appears that is proportional to the size of
the inhomogeneity $\ex-1$.
For $\rb\to \infty$ we have $\bar{f} \to 0$, so that the curvature term 
disappears and we recover the standard expansion in a spatially flat Universe.

Eq. (\ref{eom1a}) can be written as
\be
\dot{\Rb}^2(\tb,\rb)=
\frac{\ex-1+\rb^3}{\Rb}+\frac{1-\ex}{\rb},
\label{eomex} \ee
where $\tb=t H_i=\tau H_i$, $\Rb=R/r_0$, $\rb=r/r_0$, and the dot denotes a
derivative with respect to $\tb$.
Sufficiently far from the center of the configuration (large $\rb$), 
eq. (\ref{eomex}) reduces to the standard Friedmann equation 
for a homogeneous, matter dominated, flat Universe: 
$\left( \dot{\Rb}/\Rb \right)^2 \sim \Rb^{-3}$. 
Near the center of the configuration the expansion is inhomogeneous.

For the value $\ex=1$ we recover the standard homogeneous expansion.
In this case the total mass in the region $\rb <1$ could arise from 
a fluid with constant initial energy density equal to $\rho_0$.
The range $\ex > 1$ corresponds to an overdensity in the region $\rb < 1$.
This could result from a continuous distribution of matter, or from a 
singularity at the center, such as a black hole. The range 
$0\leq \ex < 1$ corresponds to an underdensity. Finally, the range 
$\ex < 0$ corresponds to a negative mass at the center.

In the case of an overdensity ($\ex>1$), there is a certain time $\tb_c$ when
$\Rb(\tb_c,1)$ is sufficiently large for the r.h.s. of eq. (\ref{eomex})
to vanish at $\rb=1$. At later times the first shell must reverse its motion
and contract. This is very similar to the phase of collapse 
in the spherical collapse model.\\\\\\

\begin{figure}[t]
\includegraphics[width=11cm, angle=270]{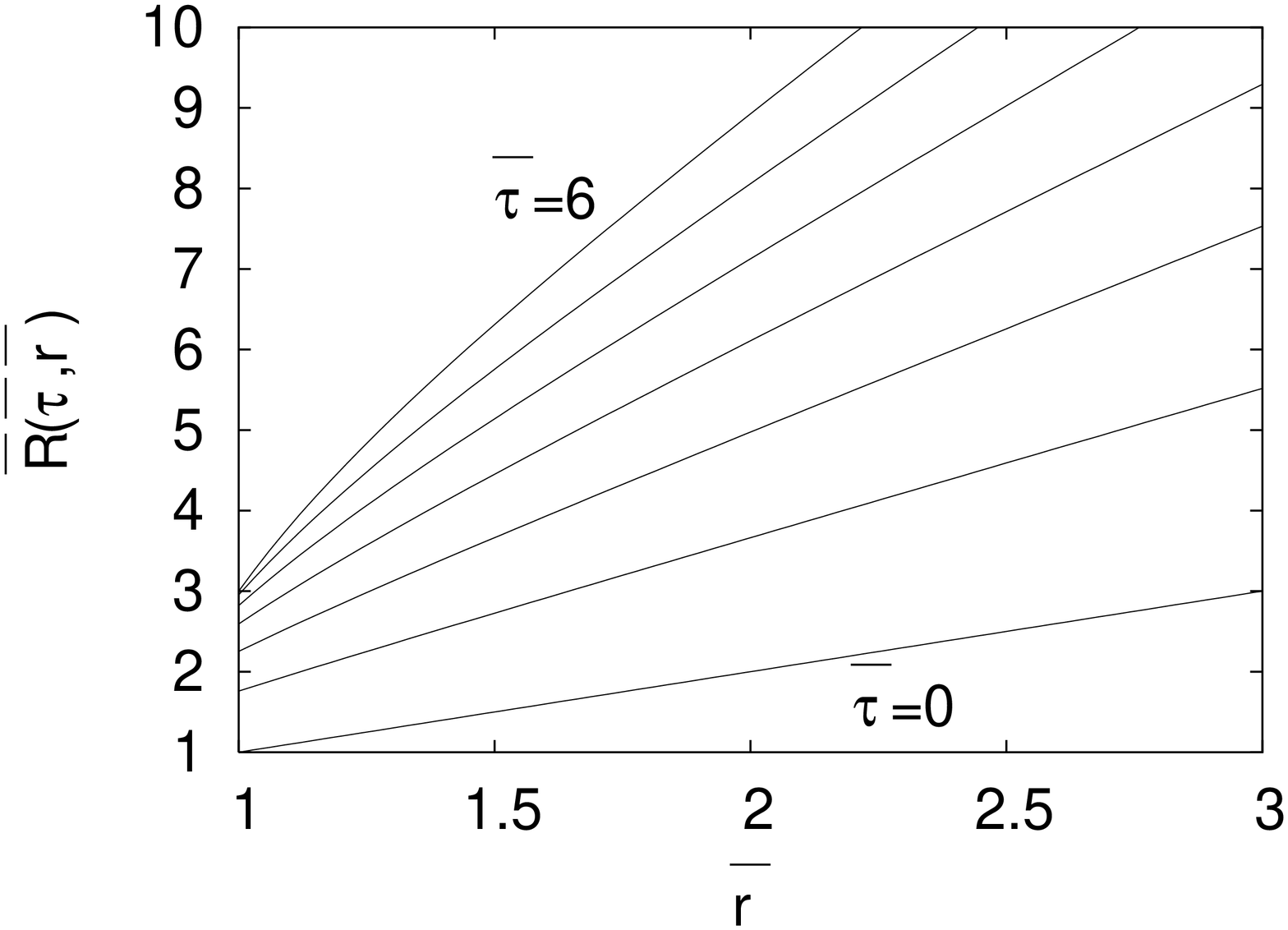}
 \caption{\it
The solution $\Rb(\tb,\rb)$ 
of eq. (\ref{eomex}) for $\ex=1.5$.
}
 \label{fig1}
 \end{figure}

\begin{figure}[t]
\includegraphics[width=11cm, angle=270]{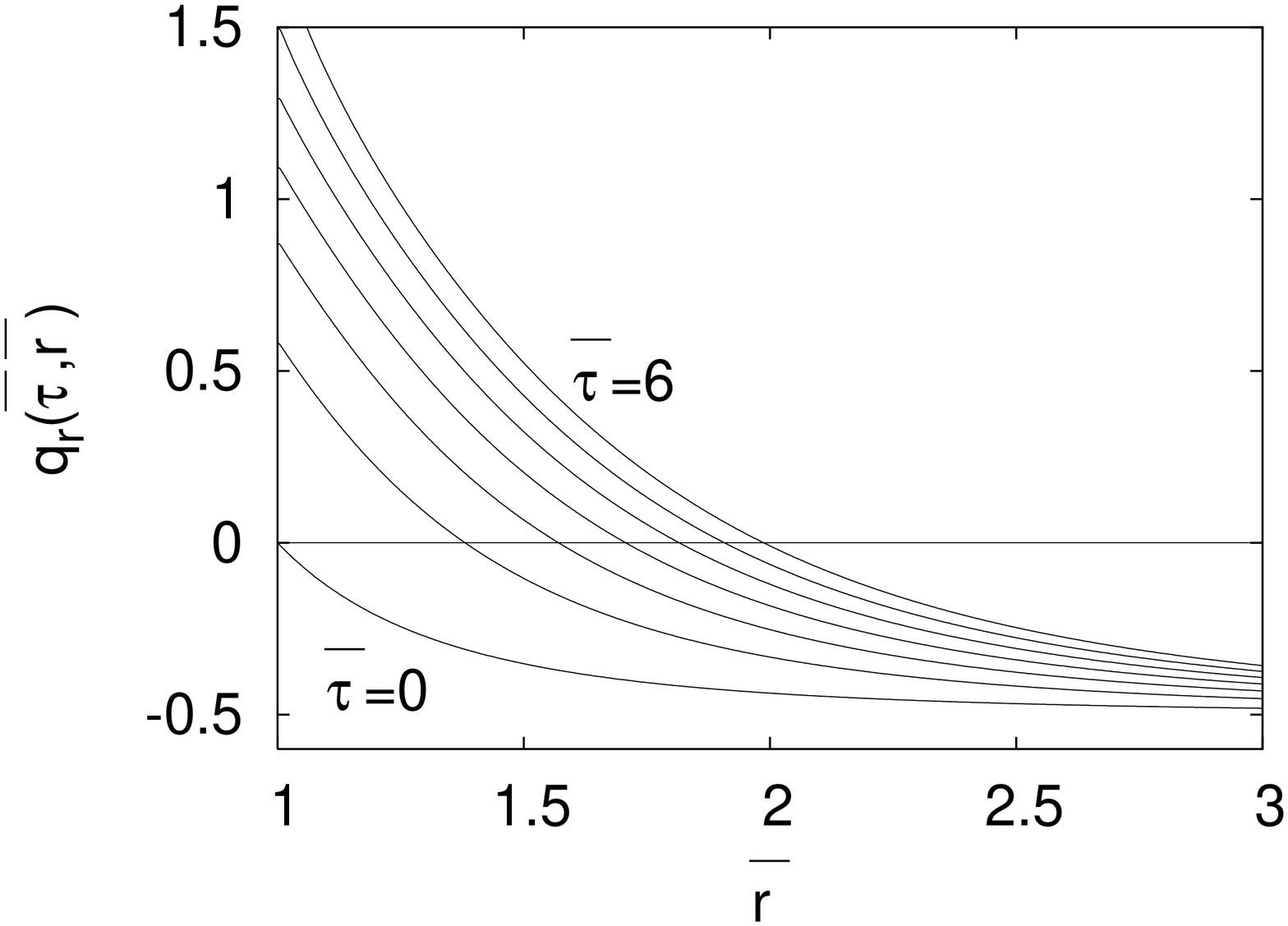}
 \caption{\it
The local acceleration parameter $q_r(\tb,\rb)$ in the radial direction for 
$\ex=1.5$.
}
 \label{fig2}
 \end{figure}

\begin{figure}[t]
\includegraphics[width=11cm, angle=270]{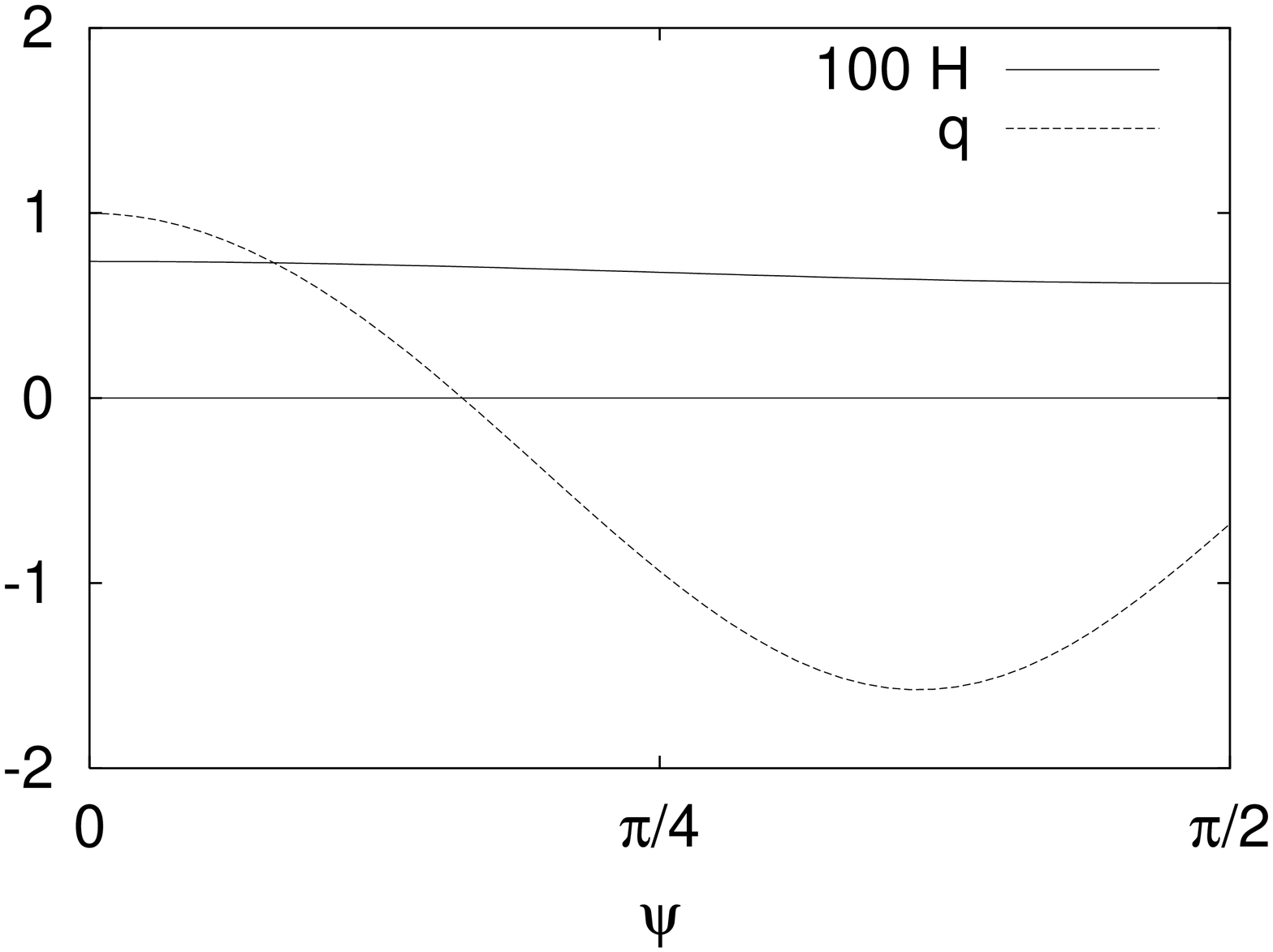}
 \caption{\it
The expansion rate $\hat{H}$ and 
the local 
acceleration parameter $\hat{q}$, as a function of the angle $\psi$ relative
to the radial direction, for $\ex=1.01$, $\rb=1$ and $\tb=100$. }
 \label{fig5}
 \end{figure}
\section{The local acceleration}
\setcounter{equation}{0}

\subsection{Positive mass}

In fig. \ref{fig1} we plot the solution of eq. (\ref{eomex})
for $\ex=1.5$, with the initial condition $\Rb(0,\rb)=\rb$.
We observe deviations from the initial linear relation between $\Rb(\tb,\rb)$
and $\rb$. These are induced by the overdensity at $\rb<1$.
In particular, the additional gravitational attraction on the
shells with $\rb \simeq 1$, resulting from the overdensity, slows down the
outward motion of these shells. At times $\tb \sim 6$ the first shell is 
beginning to reverse its motion and collapse towards the center. 
The fluid far from the overdensity is not affected significantly by it. 
Its expansion is typical of a homogeneous fluid. 

The evolution in fig. \ref{fig1} 
is very similar to the one observed in the spherical collapse
model \cite{sphc1}. An initial overdensity decouples from the homogeneous 
expansion and eventually reverses its motion and collapses. 
Our model provides additional information on the evolution of the 
material that surrounds such an overdensity. In particular, the relative
motion of two fluid shells can be determined. It is apparent from fig. 
\ref{fig1} that the $\rb$-derivative of the 
function $\Rb(\tb,\rb)$ is largest near $\rb=1$. This implies that the 
relative separation of the shells nearest to the center increases
in comparison to the unperturbed fluid at large $\rb$. 
For a comoving observer with $\rb\simeq 1$ the expansion seems faster than
in the homogeneous model. 
On the other hand, two observers with the same radial distance from the 
center, but
located at different angles, have a relative separation that tends to increase
at a slower rate than for a homogeneous fluid. 

This behaviour is reflected in the values of the acceleration parameters.
They can be expressed as
\begin{eqnarray}
q_r&=&\left(\frac{\bb}{\dot{\bb}}\right)^2
\left[\frac{\ddot{\bb}}{\bb}-\frac{1}{\bb}\left(\frac{\dot{\bb}}{\bb}\right)'
\right],
\label{qrresc} \\
q_\theta&=&\left(\frac{\Rb}{\dot{\Rb}}\right)^2
\frac{\ddot{\Rb}}{\Rb},
\label{qthresc} \end{eqnarray}
where $\bb=\Rb'/\sqrt{\bar{H}_i^{-2}+\bar{f}}$,
$\bar{f}=f/\bar{H}_i^2$, 
$\bar{H}_i=H_i r_0 $,
the dot denotes a derivative with respect to $\tb$, and the
prime a derivative with respect to $\rb$.
We use $\bar{H}_i=1$ throughout this section.
The parameter  
$q_r(\tb,\rb)$ in the radial direction is plotted in
fig. \ref{fig2}. Initially it is negative for all $\rb$.
At later times it becomes positive 
in the central region.
At all times and for large $\rb$
it approaches asymptotically the value 
$q_r =-0.5$, typical of a homogenous matter-dominated
flat Universe. 
The acceleration parameter $q_\theta(\tb,\rb)$ 
perpendicularly to the radial direction is always negative. 
For large $\rb$ it approaches the value $q_\theta=-0.5$ 
as well.

The appearance of accelerating expansion does not require $\ex \gg 1$ 
necessarily. 
In fig. \ref{fig5} we plot the expansion rate $\hat{H}$ and the acceleration
$\hat{q}$ as a function of the angle $\psi$ for a model with $\ex=1.01$.
For this we employ the expressions (\ref{hhtb}) and (\ref{qhat})--(\ref{C6}).
We consider the first shell with $\rb=1$ at a time $\tb=100$. 
The dependence of the expansion rate on the angle is mild. We can observe signs
of the slowing down of the relative
expansion perpendicularly to the radial direction: 
$H_\theta\equiv \hat{H}(\pi/2)<H_r\equiv \hat{H}(0)$. On the other hand, the 
system is far from the period of collapse, which will take place at
times $\tb ={\cal O}(10^3)$. At $\tb=100$ 
the ratio of the average density 
in the region $r<r_0$ and the density at large $r$ is
$\rho_0/\rho_\infty\simeq
\left[ (R_\infty/r_\infty)/(R_0/r_0)\right]^3\ex\simeq 1.18$. 
The local 
overdensity can still be considered 
a perturbation of the homogeneous background. 
Despite that, it induces an effective acceleration close to 
the radial direction,
for angles $\psi \lta \pi/6$.

For $0\leq \ex \leq 1$ both acceleration parameters remain negative
at all times. This means that the presence of a void cannot induce accelerating
expansion in the neighbouring regions.

\begin{figure}[t]
\includegraphics[width=11cm, angle=270]{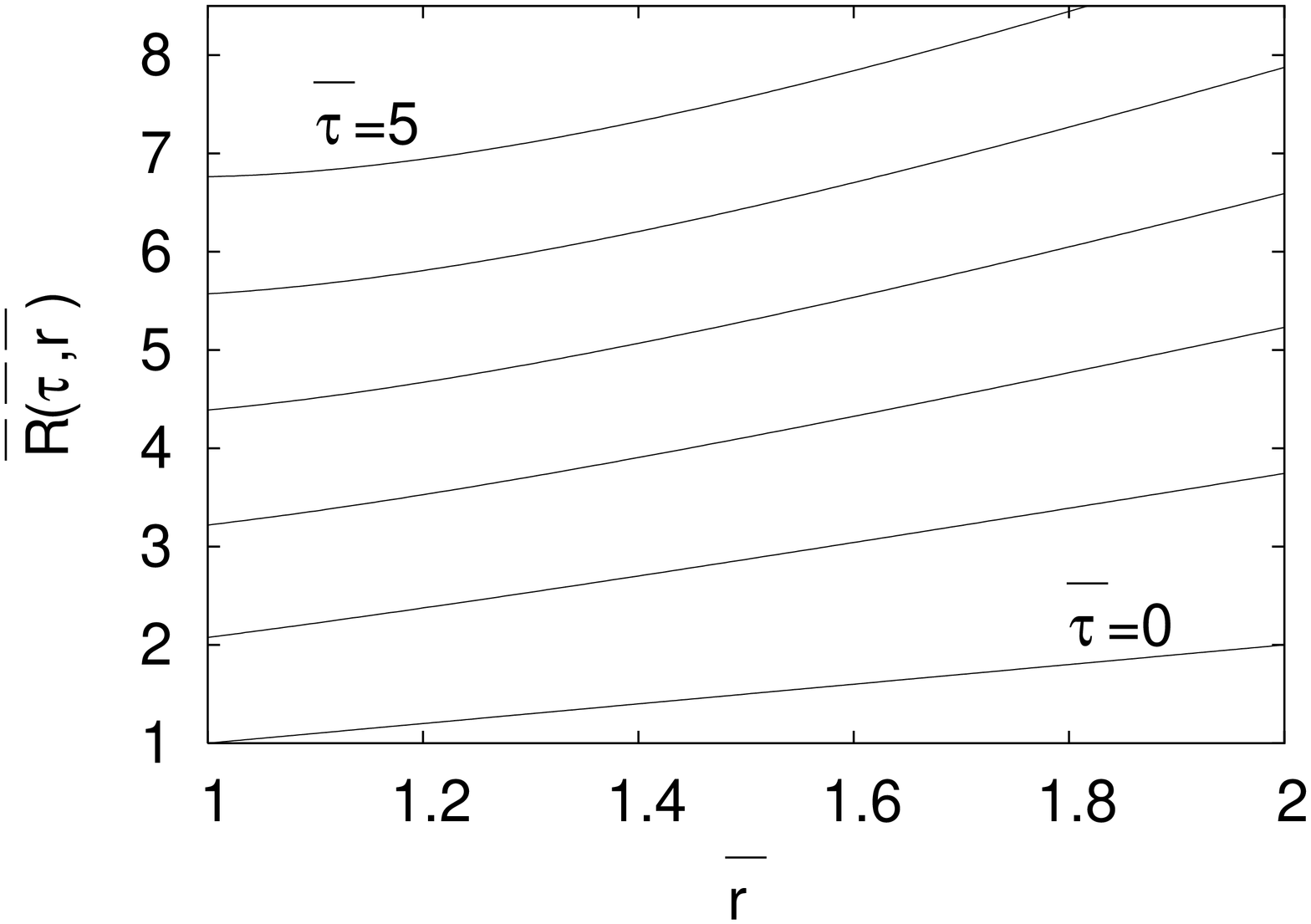}
 \caption{\it
The solution $\Rb(\tb,\rb)$ 
of eq. (\ref{eomex}) for $\ex=-0.5$.
}
 \label{fig3}
 \end{figure}

\begin{figure}[t]
\includegraphics[width=11cm, angle=270]{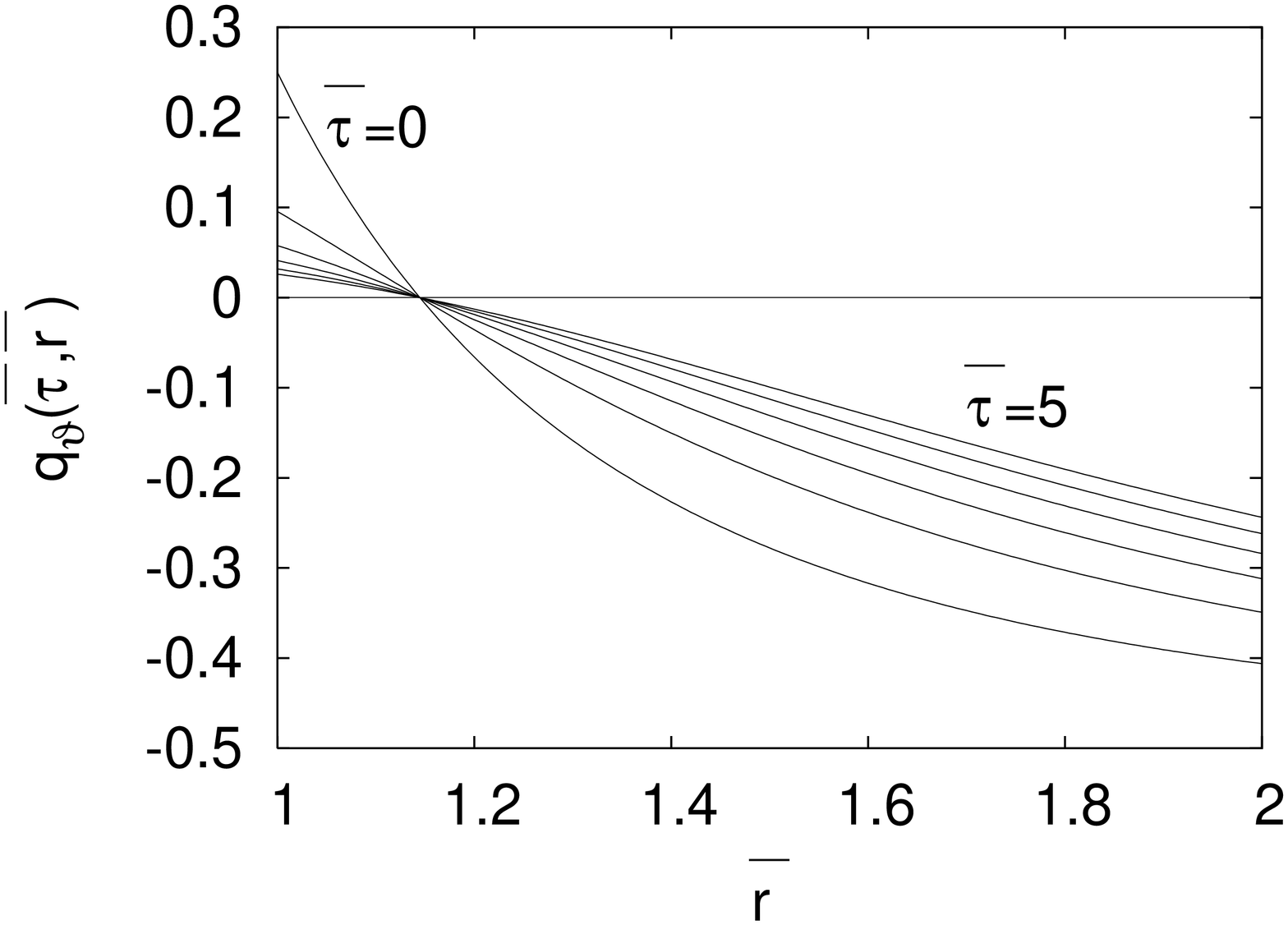}
 \caption{\it
The local acceleration parameter $q_\theta(\tb,\rb)$ perpendicularly to the 
radial direction for 
$\ex=-0.5$.
}
 \label{fig4}
 \end{figure}

\subsection{Negative mass}

In fig. \ref{fig3} we plot the solution of eq. (\ref{eomex})
for $\ex=-0.5$,
with the initial condition $\Rb(0,\rb)=\rb$.
As in fig. \ref{fig1}, we observe 
deviations from the initial linear relation between $\Rb(\tb,\rb)$
and $\rb$.
In this case, however, the shells nearest to the center are repelled strongly
by the negative mass. They expand faster than the unperturbed medium at large
$\rb$. At times $\tb\simeq 5$ the first shell catches up with the 
adjacent one (the $\rb$-derivative of $\Rb(\tb,\rb)$ vanishes at $\rb=1$)
and the phenomenon of shell crossing appears. 
This is caused by the absence of pressure in our approximation.
Our model is not adequate for the discussion of the evolution at later times.

For $0\leq \tb \lta 1$ 
the relative separation of adjacent shells near $\rb=1$ 
increases, but with a decreasing rate. 
This is reflected in the negative value of the acceleration parameter 
$q_r(\tb,\rb)$. For $\tb \gta 1$ the
relative separation decreases. As a consequence, an observer 
comoving with the fluid should perceive contraction in the radial direction.
This is confirmed by the value of the expansion rate $H_r$ of eq. (\ref{tpr1}),
which becomes negative for $\tb \gta 1$ near $\rb=1$.
On the other hand,
the relative separation of two observers with the same $\rb$, but located at
different angles, increases faster near $\rb=1$ than for large $\rb$.
The acceleration parameter 
$q_\theta(\tb,\rb)$ perpendicularly to the radial direction is plotted in
fig. \ref{fig4}. From eq. (\ref{eomex}) we find
\be
\frac{d^2\Rb}{d\tb^2} =
-\frac{\ex-1+\rb^3}{2\Rb^2}.
\label{stt} \ee
This implies that $q_\theta(\tb,\rb)$
is positive for $\rb < 1.5^{1/3}\simeq 1.15$ at all times 
for our choice of parameters.
For large $\rb$ both acceleration parameter
approach asymptotically the value 
$q =-0.5$, typical of a homogenous matter-dominated
flat Universe.

\section{The luminosity function for a central observer}
\setcounter{equation}{0}

\begin{figure}[t]
\includegraphics[width=11cm, angle=270]{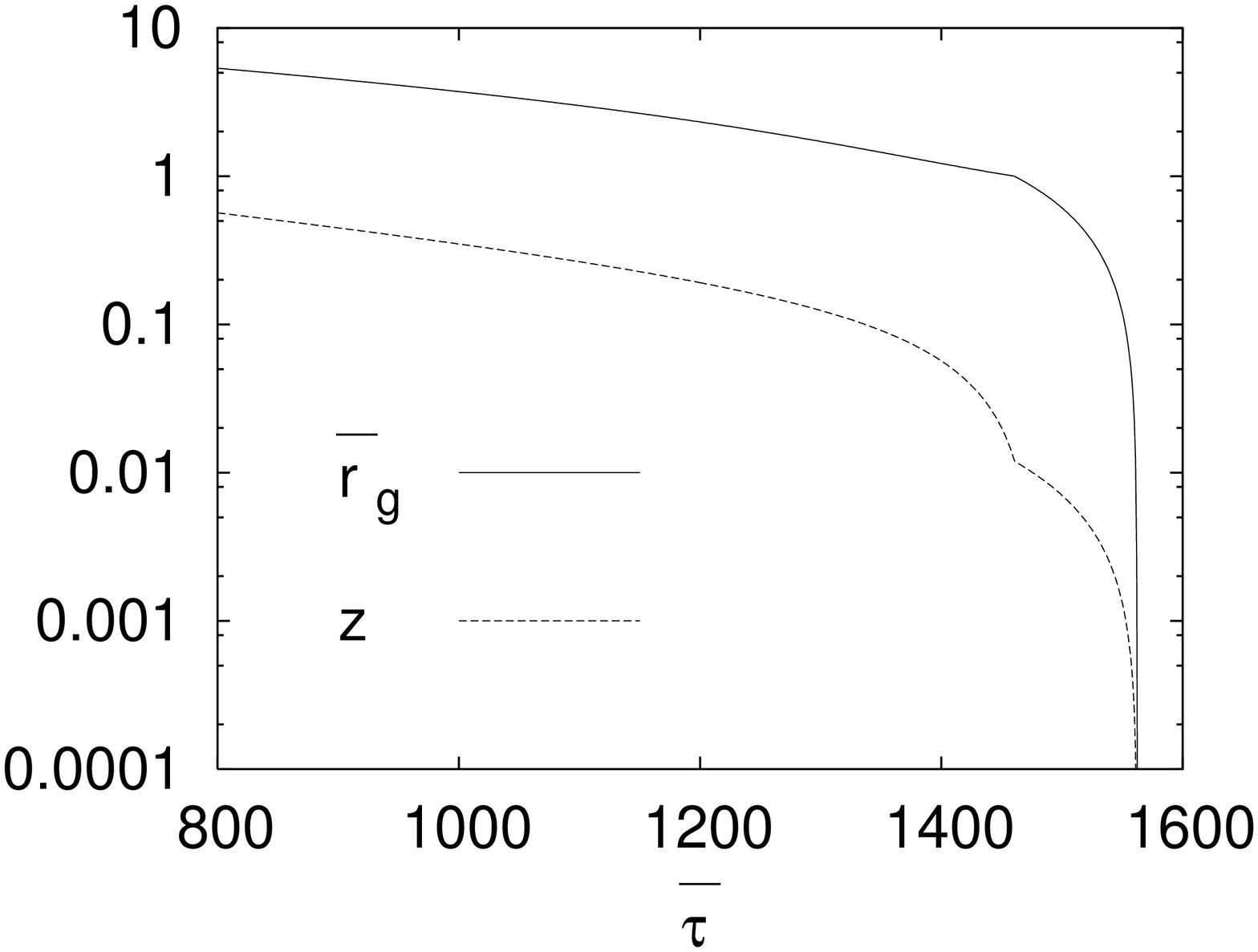}
 \caption{\it
The null geodesic $\rb_g(\tb)$ and the redshift $z(\tb)$
for a model with $\ex=1.01$, $\bar{H}_i=1$, $\tb_e=1461$, $\tb_0=1563$.
}
 \label{fig6}
 \end{figure}

\begin{figure}[t]
\includegraphics[width=11cm, angle=270]{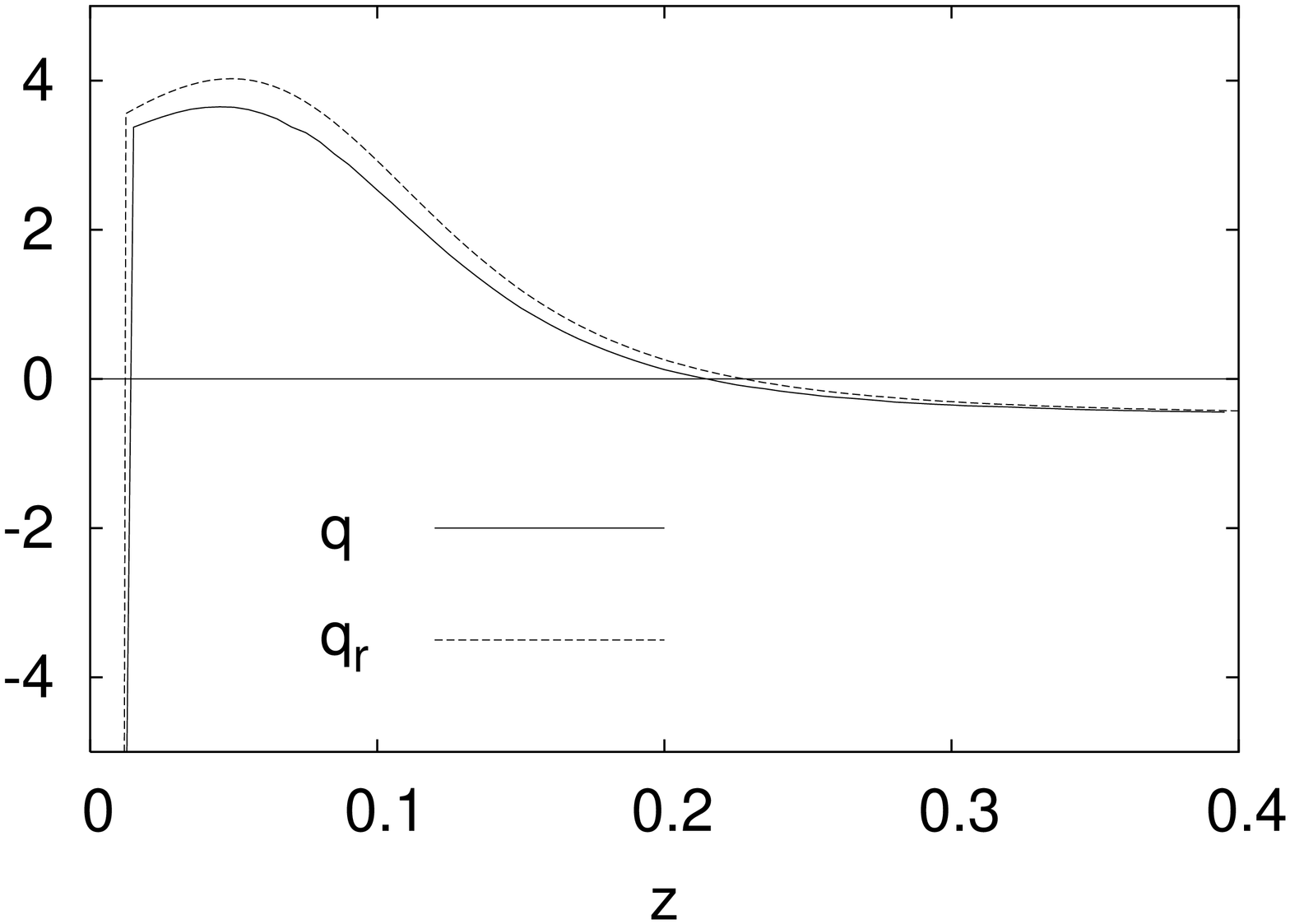}
 \caption{\it
The acceleration parameter $q$ and the local radial acceleration
$q_r$ as a function of the redshift $z$,
for a model with $\ex=1.01$, $\bar{H}_i=1$, $\tb_0=1563$, $z_0=0.012$.
}
 \label{fig7}
 \end{figure}

\begin{figure}[t]
\includegraphics[width=11cm, angle=270]{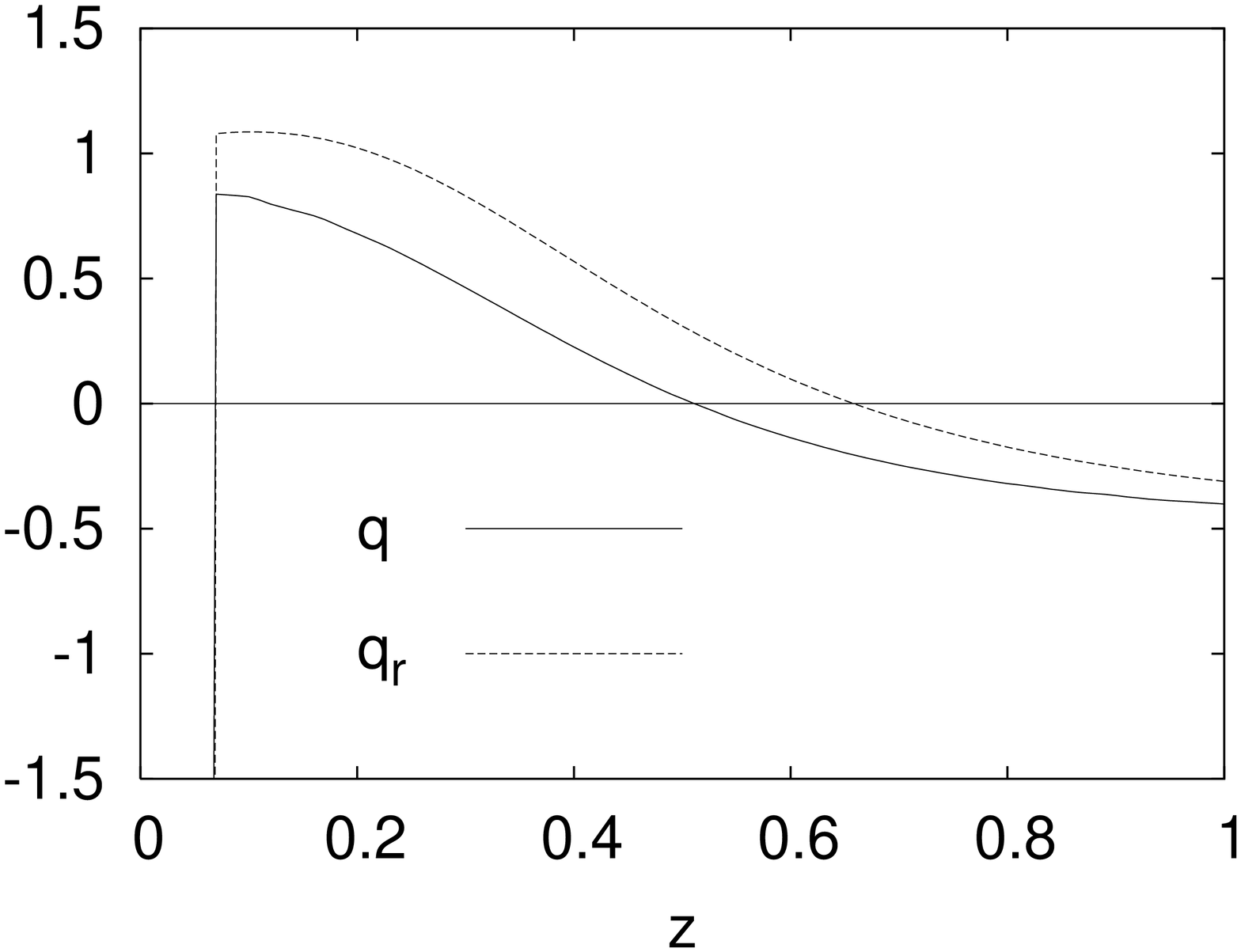}
 \caption{\it
The acceleration parameter $q$ and the local radial acceleration
$q_r$ as a function of the redshift $z$,
for a model with $\ex=1.01$, $\bar{H}_i=4$, $\tb_0=1567$, $z_0=0.067$.
}
 \label{fig8}
 \end{figure}

In this section we would like to study the issue of accelerated expansion for
an observer within the perturbation.
We assume that  the observer is located at the center of
a spherically symmetric overdensity and receives light signals from 
distances that extend beyond its surface. The assumption
of spherical symmetry makes the problem tractable. In a realistic scenario we
expect deviations from the exact symmetry, but the essense of the mechanism
should remain unaffected. On the other hand, the preferred location of the
observer is a more profound assumption. It could be argued that 
we are located near the center of a galaxy, which is a significant perturbation
in the average density. However, it is questionable if an effect at
the galactic scale can leave traces in the expansion at the level of
the horizon. On the other hand, the mechanism we consider may be relevant for
inhomogeneous cosmologies with many centers. 
The study of this situation is very difficult technically, and some kind
of averaging will have to be implemented (see e.g. \cite{averaging}).  
The necessary formalism in order to address 
this problem has not been developed yet.  

The other important feature of the scenario with 
an observer at the center of symmetry is that
all the incoming signals follow radial geodesics. The radial acceleration
can become positive, in contrast to the tangential one. In this sense, the
central location of the observer provides the optimum scenario for our 
purposes.

We also mention that 
our assumption about the location of the observer
provides automatic consistency with the 
isotropy of the cosmic microwave background. Deviations from complete isotropy,
as quantified in the difference of power between the northern and southern
hemisphere or possible allignments of the low multipoles \cite{anisotr}, 
could possibly allow for an off-center location as well. 
An alternative possibility is that the 
isotropy appears at a
sufficiently large length scale in a configuration with several centers of
symmetry. 
It remains to be seen if the
radial acceleration can be the dominant contribution for a random location of
the observer.

The simplest example of a perturbation on a homogeneous background has 
\begin{eqnarray}
\rho&=&\ex \rho_0~~~~~~~~~~~~~~{\rm for~}r\leq r_0
\label{rho1} \\
\rho&=&\rho_0~~~~~~~~~~~~~~~{\rm for~}r>r_0.
\label{rho2} 
\end{eqnarray} 
The curvature term is
\begin{eqnarray}
\bar{f}&=&(1-\ex)\rb^2~~~~~~~~~~~~~{\rm for~}\rb\leq 1
\label{ff1} \\
\bar{f}&=&\frac{1-\ex}{\rb}~~~~~~~~~~~~~~~~~{\rm for~}\rb>1.
\label{ff2} \end{eqnarray}
For $\ex>1$ the inhomogeneity is a local overdensity, for $0\leq \ex <1$ an
underdensity, while for $\ex = 1$ we have a homogeneous energy distribution.
The Friedmann equation becomes
\begin{eqnarray}
\frac{\dot{\Rb}^2}{\Rb^2}&=&
\ex\frac{\rb^3}{\Rb^3}+(1-\ex)\frac{\rb^2}{\Rb^2}
~~~~~~~~~~~{\rm for~}\rb\leq 1
\label{eom1} \\ 
\frac{\dot{\Rb}^2}{\Rb^2}&=&
\frac{\ex-1+\rb^3}{\Rb^3}+\frac{1-\ex}{\rb\Rb^2}
~~~~~~~~~{\rm for~}\rb> 1,
\label{eom2} \end{eqnarray}
where the dot denotes a partial
derivative with respect to $\tb=tH_i$. 
For $\ex=1$ we recover the standard equation for a homogeneous, 
matter dominated, flat Universe.
For $\ex>1$ the interior ($\rb\leq 1$)
expands as a homogeneous, closed Universe with spatial 
curvature proportional to $\ex-1$. The exterior ($\rb > 1$) is inhomogeneous.
For large $\rb$ we recover the Friedmann equation for a homogeneous,
matter dominated, flat Universe:
\be
\frac{\dot{\Rb}^2}{\Rb^2}=\frac{\rb^3}{\Rb^3}.
\label{eom3} \ee

We can obtain an analytical expression for the growth of an overdensity in our
model. We define the quantity 
\be
\zeta(\tb)=\frac{\Rb(\tb,\rb_\infty)/\rb_\infty}{\Rb(\tb,1)}-1,
\label{zeta} \ee
with $\rb_\infty \gg \rb_0=1$.
The ratio of the 
energy density within the perturbation to the energy density 
far from it is given by the factor 
$(1+\zeta)^3$.
Using 
eqs. (\ref{eom1}), (\ref{eom3}) we find for 
$\zeta \lta 1$ and $\tb \gta 1$ 
\be
\zeta \simeq \frac{\delta}{5} 
\left(\frac{3}{2}\tb\right)^{2/3},
\label{zetat} \ee
where $\delta=\ex-1$.
At a time $\tb_2\simeq \delta^{-3/2}$ we have $(1+\zeta)^3\simeq 2$, which
means that 
the energy density within the perturbation is double the
asymptotic value. The phase of gravitational collapse starts at a time 
$\tb_c \simeq 1.5\, \tb_2$.
The growth of a perturbation $\sim \tb^{2/3}$ is in qualitative agreement with
the behaviour predicted by the Jeans analysis for subhorizon perturbations
in a matter dominated Universe.
It is also consistent with the growth of superhorizon perturbations
in the matter dominated era. 
Our model can be viewed as an exact solution of the Einstein
equations that is consistent with the behaviour expected from perturbation
theory at all the scales of its applicability. 
This implies that we can consider values larger that one for
the dimensionless quantity $\bar{H}_i=H_i r_0$, for which the initial
perturbation extends beyond the horizon.

On radial null geodesics we have $ds^2=d\Omega^2=0$, so that the
geodesic equation becomes
\be
\frac{d\rb_g(\tb)}{d\tb}=-
\frac{\left[\bar{H}_i^{-2}+\bar{f}\left(\rb_g(\tb) 
\right)\right]^{1/2}}{\Rb'
\left(\tb,\rb_g(\tb) \right)},
\label{geod} \ee
where we have considered an incoming signal and the prime denotes a partial
derivative with respect to $\rb$. 
We first determine numerically a solution of eqs. (\ref{eom1}), (\ref{eom2})
and then use the function $\Rb(\tb,\rb)$ in order to derive a solution of
eq. (\ref{geod}). 

The redshift is given by the expression \cite{mustapha}
\be
\ln(1+z)=\int_0^{\rb_{em}} 
\frac{\dot{\Rb}'\left[\tb(\rb_g),\rb_g \right]}{
\left[1+f\left(\rb_g \right)\right]^{1/2}} d\rb_g
=
\int^{\tb_0}_{\tb_{em}} 
\frac{\dot{\Rb}'\left[\tb,\rb_g(\tb) \right]}{
{\Rb}'\left[\tb,\rb_g(\tb) \right]} d\tb.
\label{redshift} \ee
The signal
is emitted at a time $\tb_{em}$ from a point with comoving coordinate
$\rb_{em}$, and is received at the present time $\tb_0$ 
at $\rb=0$. 
The rescaled luminosity distance is \cite{celerier,humph}
\be
\bar{D}_L=(1+z)^2 \Rb(\tb_{em},\rb_{em}).
\label{luminosity} \ee
Through variation of $\rb_{em}$ we obtain the function 
$D_L(z)= r_0\,\bar{D}_L(z)$.
The expansion rate and the acceleration as a function of $z$ can be defined as
\cite{accel1}
\begin{eqnarray}
H(z)&=&\frac{1+z}{D_L'(z)}=\frac{1}{r_0}\frac{1+z}{\bar{D}_L'(z)}
\label{hubble}\\
q(z)&=&\nonumber
D_L''(z)\left[\frac{D_L'(z)}{1+z}-\frac{D_L(z)}{(1+z)^2}\right]^{-1}-1\\ 
&=&
\bar{D}_L''(z)
\left[\frac{\bar{D}_L'(z)}{1+z}-\frac{\bar{D}_L(z)}{(1+z)^2}\right]^{-1}-1,
\label{acceleration} \end{eqnarray}
where the primes denote derivatives with respect to $z$.

In fig. \ref{fig6} we plot the null geodesic $\rb_g(\tb)$ for a model with 
$\ex=1.01$ corresponding to a small initial overdensity at the level of 
$1\%$. The initial time is
$\tb_i=0.$ We take $\bar{H}_i=1$, so that initially 
the extent of the overdensity is 
comparable to the horizon: $r_0=1/H_i$. At later times the
perturbation becomes subhorizon. 
It is apparent that the form of the geodesic changes 
at a time $\tb_e = 1461$, when
the light enters the central homogeneous region of the overdensity. 
The endpoint of the geodesic is at $\rb=0$ at the present time 
$\tb_0 = 1563.$
In the same figure we also plot the redshift of photons emitted
at a certain time $\tb_{em}$ by sources located at the comoving coordinate
$\rb_g(\tb_{em})$ and observed at the present time $\tb_0$ at $\rb=0$.
The redshift of a photon emitted at the surface of the central homogeneous
region is $z_0 = 0.012.$

In fig. \ref{fig7} we depict the cosmological acceleration as a function of the
redshift. The solid line is the acceleration parameter evaluated through
eq. (\ref{acceleration}) for photons emitted at various points
along the geodesic of fig. \ref{fig6}. The dashed line is the local
radial acceleration given by eq. (\ref{qrresc}) at the point of emission. 
We observe strong deceleration for redshifts $z < z_0=0.012$. These correspond
to emission points within the central homogeneous region that expands 
like a closed Universe. On the other hand, photons emanating from immediately 
outside the central region indicate strong acceleration. The acceleration
remains positive up to redshifts $z\simeq 0.22$. Asymptotically both
acceleration parameters approach
the value $q=-0.5$, typical of a flat, matter dominated Universe. 

In fig. \ref{fig8} we repeat the calculation for a model with
the same value $\ex=1.01$, but with 
$\bar{H}_i=4$. This choice corresponds to a superhorizon initial perturbation,
as $r_0=4/H_i$. 
The present time is $\tb_0=1567$ and the redshift of the surface of the 
homogeneous region $z_0=0.067$. The central region is again strongly 
decelerating. However, there is a range of redshifts extending up to 0.5 or
0.7, for which $q$ or $q_r$, respectively, are positive. 

The direct comparison of the luminosity curves with the data from
supernova observations is not feasible within our model. 
The reason is the presence of the strong
deceleration for low redshifts.
We have seen that at times
$\tb_c \sim 1.5 \delta^{-3/2} = 1500$ 
the central region stops expanding and
reverses its motion.
On the other hand, the absence of pressure in the Tolman-Bondi fluid 
does not permit a consistent description of the collapsing phase. 
Our model combines features characterizing the cosmological expansion at
large scales and the growth of inhomogeneities at smaller scales.
Its simplicity does not allow for quantitative accuracy in its predictions.
However, our results demostrate the presence of a mechanism that could link 
the cosmological expansion with the appearance of large overdensities.

\section{Summary and conclusions}
\setcounter{equation}{0}

The notion of accelerating expansion has subtleties in inhomogeneous 
cosmology.
The usual definition of the acceleration parameter in terms of eqs. 
(\ref{theta})--(\ref{ray}) leads to eq. (\ref{rayn}) for a pressureless and
irrotational fluid. As a result the expansion is expected to be 
always decelerating if the energy density remains positive.
It is not clear, however, if the definition of 
eq. (\ref{theta}) is the most appropriate for the characterization of the
expansion. The Hubble parameter $H$ in eq. (\ref{theta})  
accounts for the volume increase 
at a given point during the expansion.
In the presence of inhomogeneities or anisotropies 
$H$ results from a certain averaging 
over the various directions. For the Tolman-Bondi metric
this is obvious from eq. (\ref{theta12}). 
On the other hand, 
the observation of light from a certain source in the sky 
provides information on the expansion along the direction to that source.
Observing sources in various directions leads to a certain averaging. In
inhomogeneous or anisotropic 
cosmologies it is not obvious that this is the same 
averaging that 
gives the volume expansion as quantified in the Hubble parameter $H$.

If there are preferred directions in the underlying geometry one can define 
several expansion parameters.
For one preferred direction, such definitions are presented in eqs. 
(\ref{tpr1}), (\ref{tpt1}), where the explicit expressions are also given for 
the case of the Tolman-Bondi metric. This metric describes the evolution of a 
spherically symmetric configuration.
For an observer located away from the 
center of the configuration, the local Hubble parameter $\Hh$ and the
acceleration parameter $\qh$ are functions of the angle between the
radial direction and that of observation.
For signals originating near the observer, 
the parameters $\Hh$ and $\qh$ can be evaluated by expanding locally the
luminosity distance of an incoming signal in powers of the redshift. 
They are given by eqs. (\ref{hhtb}) and (\ref{qhat})--(\ref{C6}), 
respectively.
In our study we considered the acceleration parameters $q_r$, $q_\theta$ 
along the radial direction and perpendicularly to it. They are given
by eqs. (\ref{qr}), (\ref{qth}), respectively.

In the first part of this work (sections 3 and 4)
we examined the effect of a mass located at the center of
a spherically symmetric 
configuration on the surrounding cosmological fluid. Our cosmological
solution is similar in spirit 
to the model of spherical collapse. 
We considered a fluid that is initially 
homogeneous and subject to uniform expansion.
We assumed that at a certain time a spherical
inhomogeneity appears at some point in space. We approximated the 
inhomogeneity as a point source located at the center of the configuration.
Its presence modifies the
local gravitational field and distorts the 
expansion of the surrounding fluid. 
For an observer located {\it away from the center} 
of such an inhomogeneity, the 
perceived local evolution is as follows:\\
a) A central overdensity leads to acceleration along the \emph{radial
direction}, and deceleration \emph{perpendicularly to it}. \\
b) A central underdensity leads to 
deceleration along and perpendicularly to the radial direction. \\
c) A negative mass at the center 
leads to deceleration along the radial
direction, and acceleration perpendicularly to it. \\
In all cases the expansion becomes typical of a homogeneous, dust-dominated
Universe far away from the inhomogeneity.

Our solution demonstrates the link between \emph{gravitational collapse near
overdensities and radial acceleration}.
Previous studies of expansion within the Tolman-Bondi model 
\cite{rasanen}--\cite{tbother} have missed this point because they
focused on the acceleration parameter
defined through the volume expansion according to 
(\ref{theta})--(\ref{ray}). This corresponds to a certain average of the
expansion rates in various directions, as is obvious from eqs. 
(\ref{tpr1})--(\ref{theta12}).
The acceleration defined in this way
remains always negative, as expected from eq. (\ref{rayn}). 
Also, in previous works the Tolman-Bondi metric has been studied in a gauge
in which the initial energy density is constant in space, while
the effect of the inhomogeneity is introduced through a function that 
determines the local Big Bang time. 
This obscures the intuition on the effect
of large concentrations of mass on the expansion.

It is apparent from our discussion that, if the observations lead to
an effective averaging of the expansion rate over various directions
similar to eq. (\ref{theta12}), deceleration should be expected. 
If, however, the averaging is modified 
the expansion could be perceived as accelerating. This would happen 
in the case of an overdensity if the 
observations are more sensitive to the expansion along the radial direction.
This is the case for
an observer located {\it at the center of an overdensity}, 
who receives light signals
only in the radial direction. In the second part part of this work 
(section 5) we 
studied the luminosity function for such an observer. The overdensity cannot
be approximated by a point source any more, as the location of the
singularity in the metric would coincide with that of the observer. For 
this reason, we assumed a continuous distribution of matter, 
that is homogeneous in the central region, falls off at larger distances, and
asymptotically becomes homogeneous with an energy density smaller than
the central one.
We were mainly interested in the light signals that  
originate in the regions outside
the central overdensity.

The form of the luminosity distance as a function of the redshift in
the above scenario provides another measure of the cosmological acceleration
through eq. (\ref{acceleration}). 
For a range of redshifts corresponding to 
photons emanating from immediately 
outside the central overdense region, we observe strong acceleration. 
The acceleration remains positive up to a certain redshift, then turns
negative, and asymptotically approaches
the value $q=-0.5$, typical of a flat, matter dominated Universe. 
For small redshifts $z \simeq 0$ we observe strong 
deceleration. This range corresponds
to photons emitted within the central homogeneous region, that evolves
like a closed Universe and eventually collapses. 
We expect this feature to be modified in more realistic collapse models.
These results are depicted in figs. \ref{fig7}, \ref{fig8}, and  
demonstrate the possible link between the \emph{growth of perturbations and 
the perceived acceleration of the expansion}.

There are several points in the above scenario
that require further work. One is the 
location of the observer, which is constrained by the 
isotropy of the cosmic microwave background. 
Exact isotropy would require the observer to be located exactly at 
the center of a spherically symmetric perturbation. Small 
deviations from complete isotropy,
such as the difference of power between the northern and southern
hemisphere or possible allignments of the low multipoles \cite{anisotr}, 
could possibly allow for an off-center location.
An alternative possibility is that the isotropy appears at a
sufficiently large length scale in a configuration with several centers of
symmetry. 
It is an interesting question whether the 
radial acceleration can be the dominant contribution to the averaged
expansion for a random location of
the observer.

The form of the expansion within a collapsing overdensity is 
another point that has to be addressed in future work. In our model 
the cosmological fluid within the overdensity 
is approximated as pressureless. 
As a result, the expansion for small redshifts is strongly 
decelerating. This makes the comparison with the supernova data problematic. 
A more sophisticated model is required in order to describe with 
quantitative accuracy both the
cosmological expansion and the collapse of an overdense region. 

Another issue concerns the existence of a characteristic scale
for the inhomogeneity. In our simple model, 
it is apparent from figs. \ref{fig7}, \ref{fig8}
that the presence of acceleration at redshifts $z \sim 0.5$ 
forces the redshift of the surface of the overdensity to values
$z_0 \sim 0.05$.
On the other hand, it is remarkable
that perturbations at the level of 10\% in the present average density 
are capable of inducing acceleration in certain directions (see fig.
\ref{fig5}). 
It is possible for such perturbations to be very extensive (of the order
of 100 Mpc or even larger) without being in
conflict with observations.

As a final remark we mention that a 
series of studies \cite{tomita} has considered the possibility that we 
are located within an underdense region of the Universe, while we receive
light signals that originate outside this region. 
The luminosity-redshift relation in this case could be
similar to the one in an accelerating homogeneous Universe.
We have verified this conclusion within our model. The expansion is 
decelerating for all redshifts, but a significant reduction in the value of
the Hubble parameter beyond a certain critical value $z_{cr}\sim 0.1$
can provide consistency with the supernova data.

\vspace {0.5cm}
\noindent{\bf Acknowledgments}\\
\noindent 
We would like to thank M. Plionis for useful discussions.
This work was supported by the research program 
``Pythagoras II'' (grant 70-03-7992) 
of the Greek Ministry of National Education, partially funded by the
European Union.

\end{document}